\documentclass[11pt]{article}

\usepackage[T1]{fontenc}
\usepackage[letterpaper,margin=1in]{geometry}
\usepackage{amsmath,amssymb}
\usepackage{graphicx}
\usepackage{booktabs}
\usepackage[hidelinks]{hyperref}
\hypersetup{
  pdftitle={Climate-Dyna Deep Hedging for XVAs: Model-Based Reinforcement Learning, Residual Climate HVA, and Hedge-Instrument Discovery},
  pdfauthor={Xiaozhen Wang and Francois Buet-Golfouse}
}

\newcommand{\E}{\mathbb{E}}

\newcommand{\Var}{\operatorname{Var}}
\newcommand{\Cov}{\operatorname{Cov}}

\title{Climate-Dyna Deep Hedging for XVAs:\\
Model-Based Reinforcement Learning, Residual Climate HVA,\\
and Hedge-Instrument Discovery}

\author{
Xiaozhen Wang\\
\small CEREMADE\\
\small Universit\'e Paris Dauphine--PSL, France\\
\small \texttt{xiaozhen.wang@dauphine.psl.eu}
\and
Francois Buet-Golfouse\\
\small AIML Global Markets\\
\small Barclays, United Kingdom\\
\small \texttt{francois.buetgolfouse@barclays.com}
}

\date{}

\begin{document}

\maketitle

\begin{abstract}
For a trading desk, residual climate hedging valuation adjustment (HVA) is the climate cost left after its inherited hedge and any admissible overlay have been taken into account; it therefore cannot be inferred from a stand-alone stress loss. We obtain this residual by comparing paired climate-on and baseline worlds and reoptimizing the overlay for each hedge universe, which also turns hedge-instrument discovery into a valuation problem: an instrument is useful to the extent that it lowers the optimized residual cost. The linear-Gaussian case has an exact finite-horizon Riccati solution; Climate-Dyna starts from that hedge and learns the remaining nonlinear correction from paired world-model rollouts, with an independent gate deciding whether to deploy the update. In a public-data-calibrated semi-synthetic EU ETS study, crediting the inherited hedge lowers the mean climate charge from $1.517$ to $0.906$, and the learned overlay lowers it to $0.831$ against a $0.821$ exact floor; residual Dyna cuts regret by $93\%$ relative to replay with one quarter as many trajectories, while adaptation from only 25 target transitions retains $60.7\%$ of the exact-assisted gain.

\end{abstract}

\noindent\textbf{Keywords:} Climate risk, XVA, Deep hedging,
Model-based Reinforcement learning, Risk-sensitive control

\section{Introduction}
\label{sec:introduction}

Climate XVA should be framed as a hedging problem. A climate transition can
move carbon prices, counterparty default probabilities, sector values, credit
spreads, rates, funding, and liquidity
\cite{dietz2016climate,battiston2017climate,capasso2020climate}. The desk
therefore needs more than a climate-stressed value: it must determine what its
existing hedge already covers, which additional instruments reduce the
residual, and what remains unhedgeable after trading costs and market-access
constraints. Existing climate adjustments price important channels
\cite{albanese2017credit,kenyon2021ccva}, but do not make the residual
exposure of a particular inherited hedge book the primary valuation object.

An ordinary credit hedge covers only a special case. Climate also creates
carbon-equivalent (XCE) cash flows, carbon-price pass-through, sector and macro
repricing, physical losses, stressed liquidity, and model-regime uncertainty.
Carbon allowances, credit default swaps (CDS), sector equities, inflation
instruments, and weather
or catastrophe contracts span different parts of this liability, often with
missing maturities and material basis risk
\cite{andersson2016hedging,engle2020hedging,bolton2021carbon,azzone2025hedging}.
The general framework therefore allows carbon, credit, sector, macro,
physical, liquidity, and model channels. Our numerical study deliberately
instantiates only the carbon, credit, sector-market, and liquidity subset.

The financial construction proceeds from desk value to residual cost.
Constant absolute risk aversion (CARA) utility under the physical measure
gives a cash-additive certainty equivalent and an entropic Bellman equation;
subtracting clean price defines
hedge-set-dependent XVA. Paired climate-on and baseline worlds then isolate
incremental climate liability using common shocks. Let
$C_{n+1}^{\mathrm{clim}}$ be the gross pre-hedge climate-on minus baseline
loss. For inherited inventory
$I_{n+1}^0$, let $\Delta H_{n+1}^{0,\mathrm{inc}}$ be its climate-on/off
incremental P\&L and let $\Delta\Gamma_{n+1}^0$ be the corresponding change
in execution, funding, margin, and liquidation costs. The liability passed to
the climate overlay is
\begin{equation}
\widehat C_{n+1}^{\mathrm{clim}}
=C_{n+1}^{\mathrm{clim}}
-(I_{n+1}^0)^\top\Delta H_{n+1}^{0,\mathrm{inc}}
+\Delta\Gamma_{n+1}^0 .
\label{eq:residual-climate-liability}
\end{equation}
Equation~\eqref{eq:residual-climate-liability} assigns the overlay only the
residual liability: it credits the inherited hedge exactly once and does not
ask the new policy to hedge the entire portfolio or the raw climate loss.

\begin{figure}[t]
    \centering
    \includegraphics[width=.88\textwidth]{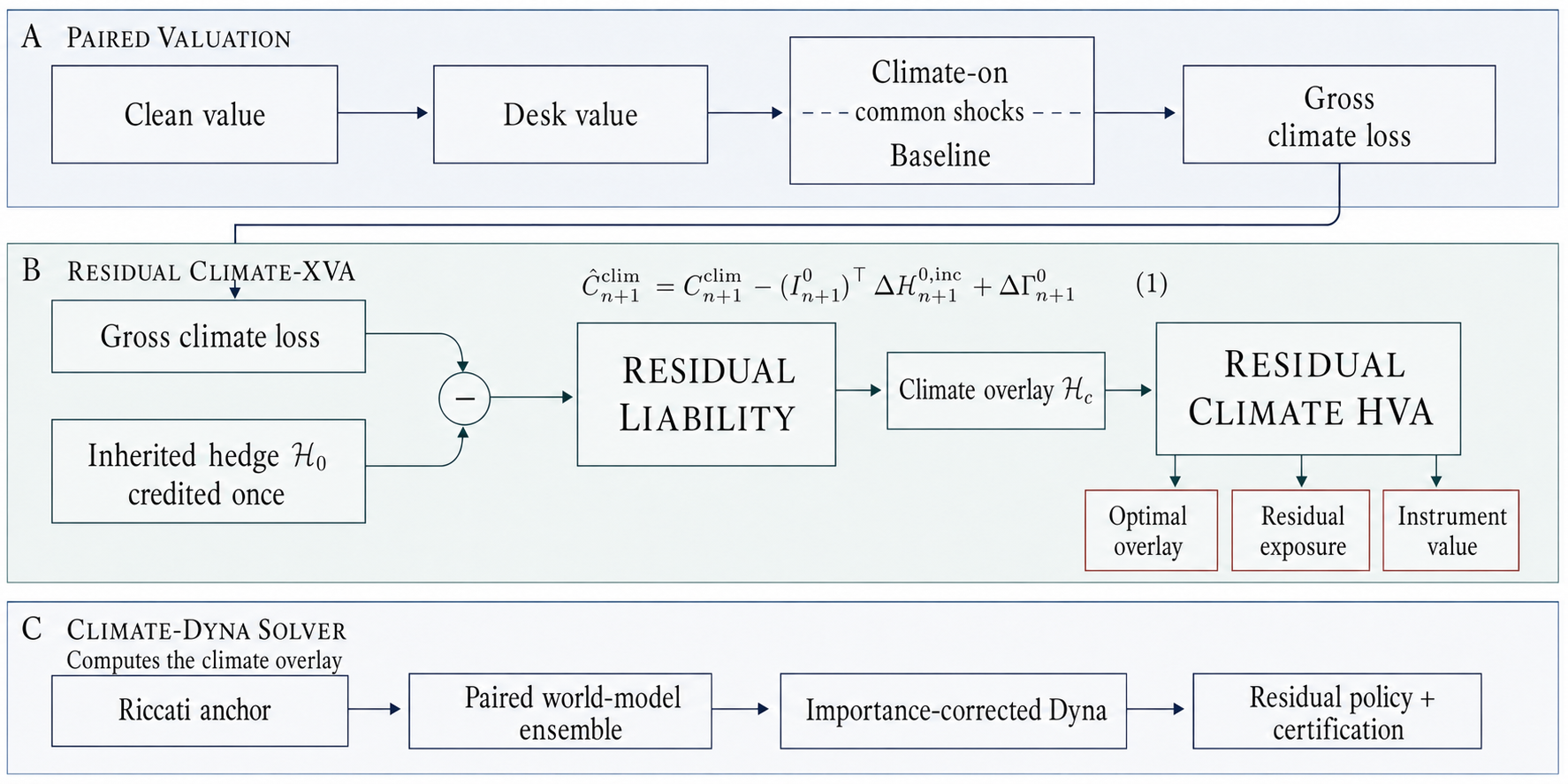}
    \caption{End-to-end framework: paired valuation isolates gross climate
    loss; inherited-hedge credit and overlay optimization yield residual
    climate HVA and instrument values; Riccati-anchored Climate-Dyna computes
    and independently certifies the overlay.}
    \label{fig:pricing-framework}
\end{figure}

Let $\Pi$ be the class of admissible overlay policies. If $\pi\in\Pi$
produces total residual loss $E_N^\pi$, the residual climate HVA is
\begin{equation}
\mathrm{HVA}_{\mathrm{clim}}^{\mathrm{res}}
:=\inf_{\pi\in\Pi}\frac1\gamma
\log\E\!\left[\exp\!\left(\gamma E_N^\pi\right)\right].
\label{eq:residual-hva}
\end{equation}
Here $\gamma>0$ measures risk aversion, and $E_N^\pi$ includes trading,
inventory, basis, and liquidation costs. The first formula states what the
overlay hedges; the second states what the desk charges.
Figure~\ref{fig:pricing-framework} summarizes this construction.

The resulting control problem is data-poor in precisely the long-horizon
regimes that matter. Historical observations do not identify repeated future
policy transitions, carbon-market completion, climate-driven default cycles,
or stressed hedge liquidity; supervisory practice consequently uses
forward-looking scenarios
\cite{ngfs2022scenarios,bcbs2021measurement}. Pure model-free learning is not
credible in this setting, while a fixed closed-form model is too rigid for
high-dimensional incomplete markets. Climate-Dyna combines the two. An exact
finite-horizon Riccati controller solves the local linear-Gaussian problem; a
constrained structural-residual world-model ensemble generates plausible
paired futures; prioritized Dyna rollouts target hedge failures; and a
risk-sensitive residual policy learns only the nonlinear correction
\cite{sutton1991dyna,chua2018deep,janner2019when}.

The experiments are a controlled instantiation rather than a claim to have
empirically completed the general framework. They test residual accounting,
Riccati recovery, model-based sample efficiency, robust policy improvement,
and hedge-set attribution in a public-data-calibrated semi-synthetic
carbon-credit-sector environment. This division is deliberate: the financial
framework specifies the broader object to be priced, whereas the experiment
provides a first computational test of selected channels.

The paper makes three contributions. First, it develops a unified residual
climate-XVA framework that connects clean value, certainty-equivalent desk
XVA, the inherited hedge, optimal climate overlays, residual climate HVA, and
instrument value. Second, it derives an exact finite-horizon Riccati overlay
and uses it as both a benchmark and an analytical prior for nonlinear deep
hedging. Third, it introduces Climate-Dyna, which combines constrained
structural-residual world modeling, importance-corrected planning,
risk-sensitive policy learning, ensemble robustness, and independent
deployment certification. The reduced study supplies proof-of-concept
evidence for this solution architecture without claiming empirical validation
of every climate transmission channel.

\section{Related Work}
\label{sec:related-work}

Climate-finance research traces policy and sector shocks through asset prices
and credit markets. Climate value-at-risk (climate VaR) and network stress
tests quantify these transmission channels
\cite{dietz2016climate,battiston2017climate,capasso2020climate}. Climate change
valuation adjustment (CCVA) and environmental credit valuation adjustment
(environmental CVA) focus more narrowly on derivatives. They allow climate
scenarios to affect counterparty default, funding, and the dependence between
default and exposure, known as wrong-way risk
\cite{kenyon2021ccva,sakuma2026environmental}. Dual Monte Carlo methods provide
upper and lower checks for controlled CVA calculations under model uncertainty
\cite{henrylabordere2016dual}. Other papers ask which traded assets can hedge
these losses. They construct portfolios from low-carbon indices, climate news,
carbon premia, and emissions networks
\cite{andersson2016hedging,engle2020hedging,bolton2021carbon,azzone2025hedging}.
We combine these two views, then measure what remains after the desk's existing
rate and credit hedge has been credited.

Reinforcement learning (RL) has also been used to price and hedge derivatives.
The Q-Learner in the Black--Scholes(--Merton) Worlds (QLBS) learns the option
value and hedge within the same risk-adjusted Q-learning problem
\cite{halperin2020qlbs}. Deep hedging replaces the linear control rule with a
neural-network trading strategy and can include convex risk measures and trading
frictions \cite{buehler2019deep}. Market impact may make a partial hedge
preferable to a full delta hedge \cite{buetgolfouse2025option}. Robust and
adversarial variants account for uncertainty in market dynamics
\cite{lutkebohmert2021robust,hirano2023adversarial}. Separate work on
multivariate time-series forecasting develops efficient neural architectures
such as Kernel-U-Net \cite{you_2024_kernel_u_net}; such models can support
transition estimation but do not themselves solve the hedging problem. The
hedging methods above generally train the whole trading strategy. We change only
the climate overlay and leave the inherited hedge fixed, so the learned policy
prices only the residual climate adjustment.

Long climate transitions are rarely observed, which makes model-based
reinforcement learning useful here. Sutton's Dyna learns from observed
transitions and from additional transitions sampled from a model
\cite{sutton1991dyna}. Probabilistic ensembles indicate where that model is
uncertain, while short rollouts limit how far its errors propagate
\cite{chua2018deep,janner2019when}. Climate-Dyna starts simulated paths from
observed states, reweights them for distribution shift, and uses a separately
fitted entropic critic to screen policy updates. Robust multi-objective
reinforcement learning changes a policy when the preference vector changes
\cite{buetgolfouse2023robust}. Our risk preference is fixed; diagnostic signals
only decide which simulated samples receive more attention.

\section{Residual Climate-XVA Framework}
\label{sec:problem}

The inherited desk hedge is taken as given. We define the climate liability
passed to an incremental overlay and the charge left after optimizing over
the instruments available to that overlay.
This separates the gross climate loss, the part already offset by the
inherited hedge, and the residual left after all admissible climate
instruments are traded.

\subsection{Climate liabilities, instruments, and paired worlds}
\label{sec:climate-system}

We work on a discrete trading grid $0=t_0<\cdots<t_N=T$. At each date $t_n$,
the desk observes the $\mathcal F_n$-measurable state
\begin{equation}
Y_n=(M_n,\chi_n,D_n,L_n,\Theta),
\label{eq:general-state}
\end{equation}
and realized desk wealth is evaluated under the physical measure $\mathbb P$.
Here $M_n$ collects traded market and counterparty-credit factors,
$\chi_n$ contains climate-policy and physical states, $D_n$ is the absorbing
default indicator, $L_n$ records liquidity and instrument availability, and
$\Theta$ contains contract and counterparty descriptors. These state variables
determine the one-period incremental climate liability, which we decompose as
\begin{equation}
C_{n+1}^{\mathrm{clim}}
=C_{n+1}^{X}+C_{n+1}^{\mathrm{credit}}
+C_{n+1}^{\mathrm{sector}}+C_{n+1}^{\mathrm{macro}}
+C_{n+1}^{\mathrm{physical}}+C_{n+1}^{\mathrm{liq}}
+C_{n+1}^{\mathrm{model}} .
\label{eq:climate-decomposition}
\end{equation}
The terms cover carbon or carbon-equivalent (XCE) cost, credit deterioration,
sector transition, macro transmission, physical damage, stressed liquidity,
and model or regime ambiguity. Natural hedges include carbon futures,
allowances, and XCE forwards; CDS and credit indices; sector equities and
commodities; rates and inflation instruments; and weather, catastrophe, and
insurance-linked contracts. Maturity mismatch, basis risk, and market closure
leave some exposure in every block.

Let $\mathcal H^0$ be the inherited rate, credit, and ordinary-XVA hedge book,
$\mathcal H^c$ a candidate climate-overlay universe, and
$\mathcal H^{\max}\subseteq\mathcal H^0\cup\mathcal H^c$ the largest
available universe after maturity, liquidity, and inventory restrictions.
Positions in $\mathcal H^0$ remain fixed; $\mathcal H^c$ is traded only
against the liability in \eqref{eq:residual-climate-liability}.

Couple the climate and baseline branches on the same one-period shock.
Conditional on the pre-trade state $Y_n$, let $\xi$ collect the structural
climate--financial parameters: scenario probabilities, pass-through and
damage coefficients, credit sensitivities, and liquidity dynamics. Both
branches use
$\varepsilon_{n+1}\sim\mathbb P(\cdot\mid\mathcal F_n)$ for uncertainty not
known at $t_n$:
\begin{equation}
Y_{n+1}^{b}=F_{\xi,n}^{b}(Y_n,\varepsilon_{n+1}),\qquad b\in\{0,1\},
\label{eq:paired-transition}
\end{equation}
where $b=1$ follows the climate-on economy and $b=0$ freezes documented
climate drifts, pass-through, credit and liquidity loadings, and regime
probabilities. Re-anchoring the baseline at observed $Y_n$ makes the branches
differ only through these mechanisms. If
$\mathcal L_{n+1}^{\mathrm{pre}}$ is economic loss before hedge P\&L, define
$C_{n+1}^{\mathrm{clim}}
:=\mathcal L_{n+1}^{\mathrm{pre}}(Y_n,Y_{n+1}^{1})
-\mathcal L_{n+1}^{\mathrm{pre}}(Y_n,Y_{n+1}^{0})$.
Holding the shock fixed removes ordinary market variation from this gross
pre-hedge increment, the first term in
\eqref{eq:residual-climate-liability}. The theory retains all channels in
\eqref{eq:climate-decomposition}; Section~\ref{sec:experiments} activates
carbon, credit, sector-market, and liquidity.

\subsection{Certainty-equivalent XVA and residual accounting}
\label{sec:certainty-residual}

We derive the desk objective from constant absolute risk aversion (CARA)
utility $U(w)=-e^{-\gamma w}$ for terminal desk wealth $w$. For a
terminal-unit payoff $Z$, the cash-additive conditional certainty equivalent
is
$\mathrm{CE}_{\gamma,n}(Z):=-\gamma^{-1}
\log\E[e^{-\gamma Z}\mid\mathcal F_n]$, where $\gamma>0$ is risk aversion.
For hedge universe $\mathcal H$, inventory $i$, and one-period terminal-unit
wealth increment $R_{n+1}^{\mathcal H}$, let
$\mathcal A_n^{\mathcal H}(y,i)$ be the feasible trade set. The desk value
satisfies
\begin{equation}
v_n^{\mathcal H}(y,i)
=\sup_{a\in\mathcal A_n^{\mathcal H}(y,i)}
\mathrm{CE}_{\gamma,n}\!\left[
R_{n+1}^{\mathcal H}
+v_{n+1}^{\mathcal H}(Y_{n+1},i+a)\right].
\label{eq:entropic-bellman}
\end{equation}
The action $a$ maximizes current wealth plus continuation value under
$\mathbb P$. Let $w_n(y)$ be the clean price under $\mathbb Q^0$, excluding
counterparty loss, funding, climate cash flows, and hedge frictions. Write
$\widetilde\Pi_{n+1}^{\mathrm{clean}}$ for the next clean contractual cash
flow and $\E_n^{\mathbb Q^0}$ for conditional expectation under
$\mathbb Q^0$:
\begin{align}
w_n(y)&=\E_n^{\mathbb Q^0}\!\left[
\widetilde\Pi_{n+1}^{\mathrm{clean}}+w_{n+1}(Y_{n+1})\right],\notag\\
u_n^{\mathcal H}(y,i)&=v_n^{\mathcal H}(y,i)-w_n(y)
\label{eq:hedge-set-xva}
\end{align}
Thus $u_n^{\mathcal H}$ is the adjustment from clean to desk value. With $0$
and $\mathrm{clim}$ denoting the baseline and climate-on economies, the
positive climate cost is
$C_{\mathrm{clim}}^{\mathcal H}
=v_0^{\mathcal H,0}-v_0^{\mathcal H,\mathrm{clim}}$.

The desk enters the overlay problem with the inherited book
$\mathcal H^0$. Let $I_{n+1}^0$ be its post-trade inventory, chosen before the
paired intervention. Its incremental climate return is
\begin{equation}
\Delta H_{n+1}^{0,\mathrm{inc}}
=\Delta H_{n+1}^{0}(Y_{n+1}^{1})
-\Delta H_{n+1}^{0}(Y_{n+1}^{0}).
\label{eq:existing-incremental-return}
\end{equation}
If $\Delta\Gamma_{n+1}^0$ is the climate-on minus climate-off change in its
execution, funding, margin, and liquidation costs, the loss handed to the
overlay is exactly $\widehat C_{n+1}^{\mathrm{clim}}$ in
\eqref{eq:residual-climate-liability}. The inherited hedge and its incremental
costs are therefore counted once, before overlay optimization.

\subsection{Optimal climate overlay and instrument value}
\label{sec:residual-hva-framework}

This subsection maps the residual liability
$\widehat C_{n+1}^{\mathrm{clim}}$ to three outputs: an optimal overlay, the
charge left after hedging, and the value of a candidate instrument. Since the
inherited book is already credited in
\eqref{eq:residual-climate-liability}, only the overlay is optimized.

\paragraph{Dynamic overlay and residual loss.}
Let $J_n$ and $\Delta H_{n+1}^c$ denote the overlay inventory and executable
return. An adapted policy chooses
$a_n=\pi_n(Y_n,J_n)$ in the feasible set
$\mathcal A_n(Y_n,J_n)$ and carries $J_{n+1}=J_n+a_n$ to $t_{n+1}$.
The feasible set and the convex costs $\Gamma_n$ and $\Phi_n$ encode liquidity,
execution, and inventory constraints. Its total loss is
\begin{align}
E_N^{\pi,\mathcal H}
=\sum_{n=0}^{N-1}\big[
&\widehat C_{n+1}^{\mathrm{clim}}
-J_{n+1}^{\top}\Delta H_{n+1}^{c}
+\Gamma_n(a_n;Y_n)\notag\\
&+\Phi_n(J_{n+1};Y_n)\big]+E_T^{\mathrm{term}} .
\label{eq:residual-error}
\end{align}
Thus the overlay P\&L is subtracted from the residual liability, while trading,
inventory, and terminal-closeout costs are added.

\paragraph{Residual climate charge.}
For hedge universe $\mathcal H$, the desk minimizes the entropic risk of this
loss over its admissible policies $\Pi(\mathcal H)$:
\begin{equation}
\mathfrak C_{\mathrm{clim}}^{\mathcal H}
=\inf_{\pi\in\Pi(\mathcal H)}
\rho_\gamma(E_N^{\pi,\mathcal H}),\qquad
\rho_\gamma(E)=\frac1\gamma\log\E[e^{\gamma E}] .
\label{eq:entropic-objective}
\end{equation}
The minimizer gives the overlay and the minimum gives its residual charge.
Because the inherited book is already credited,
$\mathfrak C_{\mathrm{clim}}^{\mathcal H^0}$ is the no-overlay benchmark.
Expanding to $\mathcal H^{\max}$ separates the hedgeable improvement
$\mathfrak C_{\mathrm{clim}}^{\mathcal H^0}
-\mathfrak C_{\mathrm{clim}}^{\mathcal H^{\max}}$
from the final residual
$\mathfrak C_{\mathrm{clim}}^{\mathcal H^{\max}}
=\mathrm{HVA}_{\mathrm{clim}}^{\mathrm{res}}$ in
\eqref{eq:residual-hva}, which retains unhedgeable basis, transaction, and
market-access losses.

\paragraph{Instrument value and span.}
Candidate instrument $j$ is valued by reoptimizing with and without it:
\begin{equation}
\Delta_j^{\mathrm{net}}(\mathcal H)
=\mathfrak C_{\mathrm{clim}}^{\mathcal H}
-\mathfrak C_{\mathrm{clim}}^{\mathcal H\cup\{j\}}
-\mathrm{Cost}_j^{\mathrm{impl}} .
\label{eq:instrument-value}
\end{equation}
Positive value means that its optimized cost reduction exceeds its fixed
onboarding cost. Reoptimization captures substitution, liquidity, and trading
costs.

For a local span diagnostic, center inherited and candidate returns
$R_{n+1}^0$ and $R_{n+1}^c$ and define
\begin{equation}
R_{n+1}^{c,\perp}
=R_{n+1}^c-\Sigma_n^{c0}(\Sigma_n^{00})^{-1}R_{n+1}^0,
\label{eq:orthogonalized-overlay-return}
\end{equation}
where $\Sigma_n^{00}=\Var_n(R_{n+1}^0)$ and
$\Sigma_n^{c0}=\Cov_n(R_{n+1}^c,R_{n+1}^0)$. This measures span beyond the
inherited book; it neither replaces reoptimization nor changes the executed
return $R^c$.

\section{Climate-Dyna Method}
\label{sec:method}

The method uses the residual liability and entropic objective from
Section~\ref{sec:problem}. It freezes the inherited hedge and computes
\eqref{eq:residual-climate-liability}. The local Gaussian Riccati action anchors
control; paired structural-model paths train nonlinear corrections. The world
model supplies plausible climate--financial transitions, while reinforcement
learning estimates only the conditional incremental trade map. Independent
rollouts decide whether a correction replaces its parent.

\subsection{Local projection and Riccati anchor}
\label{sec:riccati-anchor}

Write the overlay problem as a risk-sensitive Markov decision process (MDP):
\begin{equation}
Z_n=(n,Y_n,J_n,h_n),\qquad
a_n\in\mathcal A_n(Y_n,J_n),\qquad
r_{n+1}^{\mathrm{RL}}=-e_{n+1},
\label{eq:risk-sensitive-mdp}
\end{equation}
where $h_n$ summarizes omitted path information when $Y_n$ is not Markov and
$e_{n+1}$ is the summand in \eqref{eq:residual-error}; its negative
$r_{n+1}^{\mathrm{RL}}$ is the reinforcement-learning reward.

Before solving the dynamic problem, a one-period projection measures the
incremental span of the hedge instruments. For executable return
$R_{n+1}=\Delta H_{n+1}^c$, or $R_{n+1}=R_{n+1}^{c,\perp}$ for attribution,
set
\begin{equation}
\Sigma_n=\Var_n(R_{n+1}),\qquad
c_n=\Cov_n(R_{n+1},\widehat C_{n+1}^{\mathrm{clim}}).
\label{eq:projection-moments}
\end{equation}
Suppressing expected return and costs, the minimum-variance inventory and
remaining conditional variance are
\begin{equation}
J_{n+1}^{\mathrm{proj}}=\Sigma_n^{-1}c_n,\qquad
\sigma_{\mathrm{res},n}^2
=\Var_n(\widehat C_{n+1}^{\mathrm{clim}})
-c_n^\top\Sigma_n^{-1}c_n .
\label{eq:projection-diagnostic}
\end{equation}
Here $J_{n+1}^{\mathrm{proj}}$ is the frictionless one-period target and
$\sigma_{\mathrm{res},n}^2$ is the variance left outside that span. These are
diagnostics, not the executed dynamic strategy.
The inverse is understood on the identifiable span. For singular or nearly
collinear returns, we use the Moore--Penrose pseudoinverse, or the prespecified
ridge inverse $(\Sigma_n+\epsilon I)^{-1}$ with $\epsilon>0$, in both
quantities.

The dynamic anchor adds trading impact, inventory costs, and continuation
value. The exact anchor assumes deterministic time-dependent coefficients,
action-independent Gaussian increments, quadratic costs, and a quadratic
terminal residual. Let $\widehat C_{n+1}^{\mathrm{clim}}$ and
$\Delta H_{n+1}^c$ be jointly Gaussian with means $(\bar c_n,\mu_n)$ and
covariance terms $(\Sigma_n,c_n)$. Use the costs
$\tfrac12a^\top\Lambda_na$ for trading and
$\tfrac12x^\top Q_nx$ for post-trade inventory, and let
$(K_N,k_N,k_N^0)$ encode the quadratic terminal residual. For post-trade
inventory $x$, write
$V_{n+1}(x)=\tfrac12x^\top K_{n+1}x+k_{n+1}^\top x+k_{n+1}^0$, where
$K_{n+1},k_{n+1},k_{n+1}^0$ are its quadratic, linear, and constant
coefficients, and define
\begin{equation}
A_n=\Lambda_n+K_{n+1}+Q_n+\gamma\Sigma_n,\qquad
d_n=\mu_n+\gamma c_n-k_{n+1}.
\label{eq:riccati-terms}
\end{equation}
The Gaussian log moment-generating function and quadratic costs give
\begin{align}
J_{n+1}^{R,\star}(J_n)
&=A_n^{-1}(\Lambda_nJ_n+d_n),\label{eq:riccati-policy}\\
K_n&=\Lambda_n-\Lambda_nA_n^{-1}\Lambda_n,\qquad
k_n=-\Lambda_nA_n^{-1}d_n,\notag\\
k_n^0&=\bar c_n+\frac{\gamma}{2}
\Var_n(\widehat C_{n+1}^{\mathrm{clim}})
+k_{n+1}^0-\frac12d_n^\top A_n^{-1}d_n .
\label{eq:riccati-recursion}
\end{align}

If the joint covariance is positive semidefinite, $A_n\succ0$ at every
date, and the action set is unconstrained, or equivalently the feasibility
constraints are inactive, backward induction and completion of squares show
that the value function remains quadratic and
\eqref{eq:riccati-policy} is the unique optimal post-trade inventory in this
anchor. Define
$a_n^R(Y_n,J_n):=J_{n+1}^{R,\star}(J_n)-J_n$, with state dependence through
the conditional moments and costs. When constraints bind, we project this
action onto $\mathcal A_n(Y_n,J_n)$ and use it as a local parent rather than
claim exact optimality. The superscript $R$ denotes Riccati.

The scalar constant-coefficient case also has an exact finite-horizon closed
form. Let $q=Q+\gamma\sigma_H^2$, where $\sigma_H^2$ is hedge-return variance,
and take $K_N=0$. The recursion is the M\"obius iteration
$K_n=F(K_{n+1})$ with $F(k)=\Lambda(k+q)/(\Lambda+k+q)$. For $q>0$, define
\begin{equation}
\begin{aligned}
\kappa_\pm&=\frac{-q\pm\sqrt{q^2+4\Lambda q}}{2},&
\rho&=\frac{\Lambda-\kappa_+}{\Lambda-\kappa_-},\\
K_n&=\frac{\kappa_+-r_m\kappa_-}{1-r_m},&
r_m&=\rho^m\frac{\kappa_+}{\kappa_-},\qquad m=N-n .
\end{aligned}
\label{eq:scalar-riccati-closed-form}
\end{equation}
If $q=0$, then $K_n=0$. Thus the solution is exact on the trading grid:
$q$ fixes the long-run target curvature, while $\Lambda$ governs the speed of
adjustment toward it.

The one-period frictionless limit separates hedging from speculative demand.
If $\Lambda_n=K_{n+1}=Q_n=k_{n+1}=0$, then
\begin{equation}
J_{n+1}^{R,\star}
=\Sigma_n^{-1}c_n+\gamma^{-1}\Sigma_n^{-1}\mu_n .
\label{eq:hedge-risk-premium}
\end{equation}
The first term is the regression hedge for the residual climate liability; the
second is risk-premium demand. A pure climate-overlay mandate sets $\mu_n=0$,
or projects expected return out before estimation. Trading impact, inventory
costs, and continuation value turn this static decomposition into
\eqref{eq:riccati-policy}. Deterministic carry terms can be added to $k_n^0$;
they change desk value but not the optimal action.

\subsection{Paired world model and Dyna planning}
\label{sec:world-dyna}

Historical data identify current markets and short-run dependence, but not
future climate-policy or physical regimes. We therefore keep the paired
structural transition and learn only corrections around it. Let $\mathcal T$
map constrained continuous state coordinates to an unconstrained
representation. For ensemble member $k$ and branch $b\in\{0,1\}$, set
\begin{align}
X_{n+1}^{b,k}
={}&F_{\xi_k,n}^{\mathrm{str},b}(Y_n,\varepsilon_{n+1})
+B_n^{\mathrm{sh}}G_{\psi_k,n}^{\mathrm{sh}}
(Y_n,\varepsilon_{n+1})\notag\\
&+bB_n^{\mathrm{cl}}G_{\psi_k,n}^{\mathrm{cl}}
(Y_n,\varepsilon_{n+1}),\qquad
Y_{n+1}^{b,k}=\mathcal T^{-1}(X_{n+1}^{b,k}).
\label{eq:paired-world-model}
\end{align}
Here $X$ is the unconstrained next state and
$F^{\mathrm{str}}_{\xi_k}$ is the structural transition.
$G^{\mathrm{sh}}_{\psi_k}$ corrects errors shared by both branches, whereas
$G^{\mathrm{cl}}_{\psi_k}$ corrects only the climate mechanism;
$B_n^{\mathrm{sh}}$ and $B_n^{\mathrm{cl}}$ select the affected coordinates.
The factor $b$ removes the climate correction from the baseline branch.
Masked-softmax regime rows, nonnegative intensities, and absorbing default
enforce the state constraints. At update $r$, the ensemble
$\mathcal M_r=\{(\xi_k,\psi_k)\}_{k=1}^{K_{\mathrm{ens}}}$ represents
uncertainty in both the structural and correction parameters.

At each update, the model is fitted without discarding this structure. Each
member uses observed market, disclosure, liquidity, and desk transitions,
together with paired scenario anchors:
\begin{equation}
\mathcal L_{\mathrm{wm}}^{(k)}
=\mathcal L_{\mathrm{obs}}^{(k)}
+\lambda_{\mathrm{pair}}\mathcal L_{\mathrm{pair}}^{(k)}
+\lambda_{\mathrm{con}}\mathcal R_{\mathrm{con}}^{(k)}
+\lambda_{\mathrm{res}}\mathcal R_{\mathrm{res}}^{(k)} .
\label{eq:world-model-loss}
\end{equation}
$\mathcal L_{\mathrm{obs}}^{(k)}$ fits observed transitions and
$\mathcal L_{\mathrm{pair}}^{(k)}$ fits the paired anchors.
$\mathcal R_{\mathrm{con}}^{(k)}$ penalizes state-constraint violations,
no-arbitrage failures in traded prices, and inconsistencies with XCE
term-sheet accounting; $\mathcal R_{\mathrm{res}}^{(k)}$ keeps corrections
small. All weights are nonnegative. Data may correct the structural model, but
cannot override its financial constraints.

The resulting training set still contains no historical labels for optimal
climate trades. Observed desk segments identify current markets, short-run
dependence, liquidity, execution costs, and realized P\&L, but cannot label
hedges for policy transitions or physical regimes that have not occurred.

At update $r$, new observations first refit $\mathcal M_r$. Hybrid rollouts
extend observed desk segments; synthetic rollouts cover longer-horizon
regimes. The policy is trained on both. Every rollout freezes the inherited
hedge, constructs \eqref{eq:residual-climate-liability}, and changes only the
climate overlay. Large residual losses show where further planning is useful.

Uniform rollouts mostly revisit states in which the hedge already works. For a
candidate trajectory $\tau$ drawn under the unprioritized law $p_{0,k}$, we
instead sample from
\begin{equation}
q_{\beta,k}(\tau)\propto p_{0,k}(\tau)
\exp\!\left(\beta_RS_R+\beta_US_U+\beta_BS_B+\beta_LS_L\right),
\label{eq:planning-priority}
\end{equation}
Here $\beta_R,\beta_U,\beta_B,\beta_L\geq0$ are priority weights for residual
loss $S_R$, model disagreement $S_U$, basis risk $S_B$, and liquidity stress
$S_L$.

This prioritization changes the sampling law, not the target objective. For
policy $\pi_\phi$, let $E_N^{m,\pi_\phi,k}$ be \eqref{eq:residual-error} on
path $\tau_m$ under member $k$. Importance weights return the prioritized
sample to the original law:
\begin{equation}
\mathcal L_{\mathrm{plan},k}(\phi)
=\frac1\gamma\log
\frac{\sum_m w_{m,k}\exp\{\gamma E_N^{m,\pi_\phi,k}\}}
{\sum_m w_{m,k}},\qquad
w_{m,k}=\frac{p_{0,k}(\tau_m)}{q_{\beta,k}(\tau_m)} .
\label{eq:importance-entropic-loss}
\end{equation}
With common support and a finite exponential moment, this is consistent for
the objective under $p_{0,k}$. If the weights are omitted, the result is a
different, deliberately stress-tilted objective.

\subsection{Residual policy, robustness, and selection}
\label{sec:risk-learning}

Let $\Delta a_\phi$ be the neural correction to the Riccati action. The
executable policy is
\begin{equation}
\pi_\phi(n,Y_n,J_n,h_n)
=\operatorname{Proj}_{\mathcal A_n}\!\left[
a_n^R(Y_n,J_n)+\Delta a_\phi(n,Y_n,J_n,h_n)\right].
\label{eq:residual-policy}
\end{equation}
Here $\operatorname{Proj}_{\mathcal A_n}$ enforces feasible actions. The
residual network learns effects left outside the local Gaussian model. Its
output is partitioned into the carbon/XCE, credit, sector, macro, and physical
instrument groups of Section~\ref{sec:residual-hva-framework}; availability
masks set closed or missing instruments to zero. Group regularization aids
interpretation but does not replace bid--ask, impact, margin, roll, maturity,
or inventory costs already charged in \eqref{eq:residual-error}.

There are two complementary policy updates. When rollouts are differentiable,
the actor is trained directly. For model member $k$ and $M$ simulated paths,
the residual policy minimizes
\begin{align}
\widehat{\mathcal L}_k(\phi)
&=\frac1\gamma\log\!\left(
\frac1M\sum_{m=1}^M
\exp\{\gamma E_N^{m,\pi_\phi,k}\}\right),\notag\\
\nabla_\phi\widehat{\mathcal L}_k
&=\sum_m\omega_m\nabla_\phi E_N^{m,\pi_\phi,k}.
\label{eq:empirical-entropic-loss}
\end{align}
where
$\omega_m=\exp\{\gamma E_N^{m,\pi_\phi,k}\}/
\sum_{m'=1}^M\exp\{\gamma E_N^{m',\pi_\phi,k}\}$.
The tilted gradient emphasizes paths with large residual loss.

A critic-based update is used when target-market data are too sparse for stable
whole-path gradients. Let $f$ index a held-out fold, $s$ a sample with weight
$w_s$, $y$ the current state, $J$ current inventory, and $x$ post-trade
inventory. The continuation estimate $\widehat V^{k,f}$ excludes fold $f$;
$g_n(y,J,x)$ is current cost; and $\widehat C^{s,k},\Delta H^{c,s,k}$ are the
sampled residual liability and hedge return, with $Y_{n+1}^s$ their next state:
\begin{align}
\ell_{n+1}^{s,k}(x)
&=\widehat C_{n+1}^{s,k}
-x^\top\Delta H_{n+1}^{c,s,k}
+\widehat V_{n+1}^{k,f}(Y_{n+1}^s,x),\notag\\
\widehat Q_n^{k,f}(y,J,x)
&=g_n(y,J,x)+\frac1\gamma\log
\frac{\sum_s w_s\exp\{\gamma\ell_{n+1}^{s,k}(x)\}}
{\sum_s w_s}.
\label{eq:backward-entropic-critic}
\end{align}
Here $\ell^{s,k}$ is loss plus continuation and $\widehat Q^{k,f}$ its
entropic fold value. For a nonempty compact feasible set, the critic proposes
the post-trade inventory
\begin{equation}
\begin{aligned}
x_n^{\star,k,f}(y,J)
&\in\arg\min_x\ \widehat Q_n^{k,f}(y,J,x),\\
&\text{subject to }x-J\in\mathcal A_n(y,J),\\
a_n^{\star,k,f}&=x_n^{\star,k,f}-J .
\end{aligned}
\label{eq:critic-policy-improvement}
\end{equation}
The residual actor distills
$a_n^{\star,k,f}-a_n^R(y,J)$ rather than relearning the entire hedge. Because
fold $f$ is excluded from the continuation fit, its targets provide an
out-of-fold policy-improvement signal. Thus
\eqref{eq:empirical-entropic-loss} is the direct pathwise route, whereas
\eqref{eq:backward-entropic-critic}--\eqref{eq:critic-policy-improvement} give
the critic-based route used for scarce-data adaptation.

Let $\mathcal L_{\mathrm{obs},k}(\phi)$ be the observed-seeded loss and
$\mathcal L_{\mathrm{plan},k}(\phi)$ the corrected loss in
\eqref{eq:importance-entropic-loss}. For $\alpha\in[0,1]$, set
$\mathcal L_k(\phi)=\alpha\mathcal L_{\mathrm{obs},k}(\phi)
+(1-\alpha)\mathcal L_{\mathrm{plan},k}(\phi)$ and
$\bar{\mathcal L}(\phi)=K_{\mathrm{ens}}^{-1}\sum_k\mathcal L_k(\phi)$.
The end-to-end objective is
\begin{equation}
\min_\phi\ 
\bar{\mathcal L}(\phi)
+\eta_{\mathcal M}\operatorname{SD}_k(\mathcal L_k(\phi))
+\lambda_{\mathrm{dup}}\mathcal R_{\mathrm{dup}}(\phi)
+\lambda_{\mathrm{grp}}\mathcal R_{\mathrm{grp}}(\phi).
\label{eq:climate-dyna-objective}
\end{equation}
Here $\operatorname{SD}_k$ is ensemble standard deviation; the three
coefficients are nonnegative weights; $\mathcal R_{\mathrm{dup}}$ penalizes
exposure spanned by $\mathcal H^0$; and $\mathcal R_{\mathrm{grp}}$ imposes
instrument-group sparsity. Setting $\eta_{\mathcal M}=0$ gives
scenario-mixture training; increasing it penalizes sensitivity to the
world-model member and approximates a more conservative hedge.

Policy fitting and deployment are separate. On independent whole-policy
rollouts, define the gate difference
$D^{\mathrm{gate}}=\mathcal L_{\mathrm{cand}}-\mathcal L_{\mathrm{parent}}$.
A candidate replaces its parent only if the upper confidence bound for
$\E[D^{\mathrm{gate}}]$ is nonpositive under both the reference model and the
stressed-liquidity model. Common-path reoptimization with and without
instrument $j$ estimates $\Delta_j^{\mathrm{net}}$ in
\eqref{eq:instrument-value}. To account for substitution between instruments,
we also use the Shapley attribution
$\operatorname{Sh}_j=\E_\sigma[
\mathfrak C_{\mathrm{clim}}^{S_\sigma(j)}
-\mathfrak C_{\mathrm{clim}}^{S_\sigma(j)\cup\{j\}}]$, where $\sigma$ is a
random instrument ordering and $S_\sigma(j)$ contains the instruments preceding
$j$. Reports include the overlay by instrument and date, the residual climate
HVA and its distribution, uncertainty bands, and net instrument values. The
learned policy therefore does not estimate a causal climate law. It learns
which incremental trades best reduce the residual liability under the stated
world-model ensemble and which exposure remains unspanned.

\section{Empirical Evaluation}
\label{sec:experiments}

\subsection{Data, Paired Environment, and Protocol}
\label{sec:data-protocol}

\textit{Scope and data.}
We evaluate the carbon, credit, sector-market, and liquidity channels in
\eqref{eq:climate-decomposition}. The tests cover residual accounting, the
Riccati benchmark, Dyna training, adaptation, and instrument attribution. We
do not claim evidence for XCE term sheets, physical-risk hedges, or live-desk
use. The environment is a weekly semi-synthetic EU Emissions Trading System
(EU ETS) model calibrated to public data from 2012--2025. European Energy
Exchange (EEX) auctions provide allowance prices and liquidity observations
\cite{eex2026auctions}. Volume-weighted price calibrates $P_n^X$; inverse cover,
bid volume, and participation enter $L_n$. We also use the EXH9 utilities ETF,
the ECB three-month spot rate, the bond component of CISS, EU Transaction Log
emissions, and EBA PD/LGD ranges
\cite{ishares2026exh9,ecb2026yield,ecb2026ciss,
eea2026ets,eba2026dashboard}. Because auction prices are not futures
quotes, we add basis and roll noise to the tradable carbon return.

\textit{Paired environment.}
Only the climate-on transition is observed, so the climate-off branch is
constructed rather than estimated. It starts from the same state and uses the
same innovations, but freezes climate drift, pass-through, credit and liquidity
loadings, and regime probabilities at pre-Market Stability Reserve estimates.
The chronological split is 2012--2018 for training, 2019--2020 for validation,
and 2021--2025 for out-of-distribution testing. Each episode lasts 26 weeks and
uses a balanced, carbon-heavy, or credit-heavy book. Inherited rate and CDS
positions are fitted on the training period and held fixed thereafter.

\textit{Validation and comparison.}
Before comparing policies, we run three implementation checks. A stable vector
autoregression is fitted to pre-Market Stability Reserve innovations and used
in both branches. The Riccati recursion is checked against an independent
quadratic program under spread and inventory stresses. We also verify every
liability and overlay identity path by path. Exact dynamic programming is an
evaluation oracle, not a training input. The component study uses 20 held-out
datasets with 250 observed transitions and adds ensemble fitting, Riccati
residualization, observed seeding, priority, importance correction, and
robustness in turn. Under a matched budget, we compare replay, model-based
scratch, residual Dyna, and robust Dyna after 15/30/60/120 updates, with 200
gradient paths and 500 candidate paths per update.

\textit{Adaptation and certification.}
For adaptation, we give a source-market policy only 25 transitions from a
stable, moderate, or severe target law. We compare a cross-fitted critic and
its independent Monte Carlo gate with posterior Riccati, pessimistic fitted
Q-iteration (FQI), and an exact-assisted reference. The scalable version uses
eight folds, four world models, and 128 samples per row. Exact critic and policy
values remain hidden until all decisions are fixed. The deployable gate uses
only candidate-minus-parent upper confidence bounds on independent reference
and stressed-liquidity rollouts (Section~\ref{sec:risk-learning}). Exact regret,
improvement over Riccati and FQI, retained exact-assisted gain, and
deteriorations above $0.001$ are post-selection outcomes on 20 unused seeds,
not gate inputs; uncertainty uses 20,000 paired bootstrap resamples. Finally,
we reoptimize with no overlay, EUA only, utilities only, and both instruments
to obtain the instrument attribution.

\begin{figure}[!t]
  \centering
  \includegraphics[width=.96\textwidth]{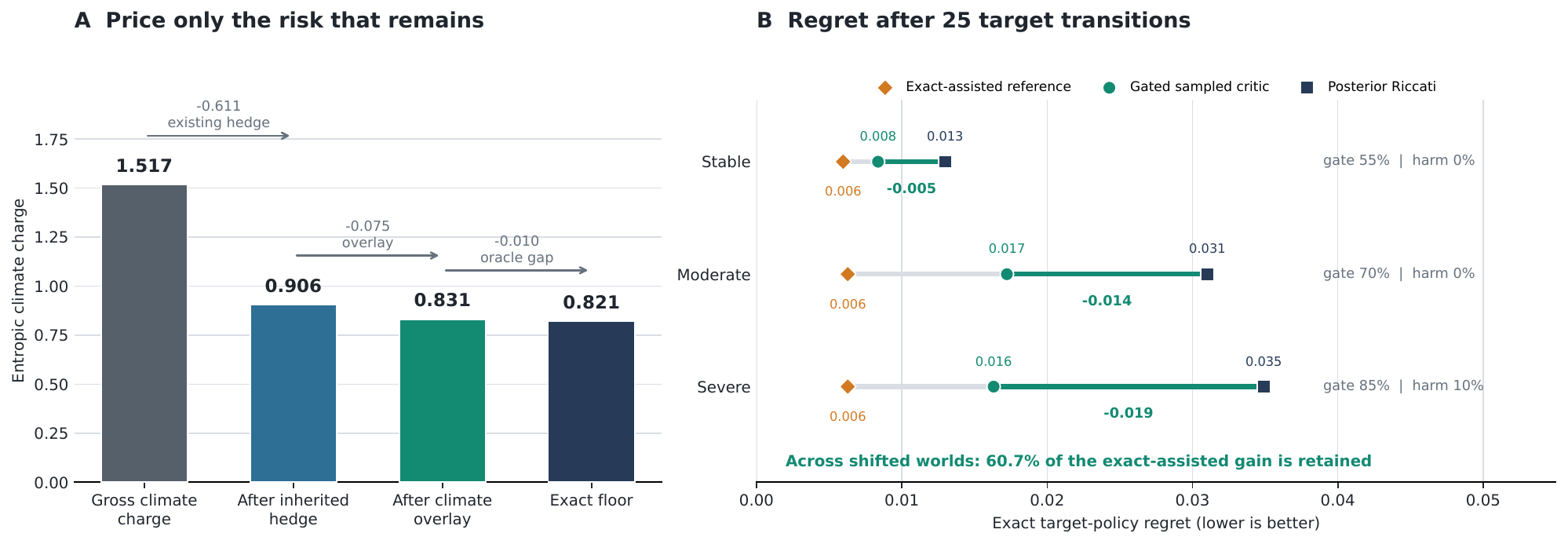}
  \caption{Financial and control residuals. (A) The inherited hedge is credited
  before the climate overlay is valued. (B) Posterior Riccati, the gated sampled
  critic, and the exact-assisted reference are compared after 25 target
  transitions; movement to the left means lower regret. The panels use
  different numerical scales.}
  \label{fig:headline-story}
\end{figure}

\begin{table}[!t]
\caption{Adaptation after 25 target-market transitions on 20 unused seeds.
Reported regrets are exact evaluation outcomes (lower is better).
$\Delta_{\mathrm{R}}$ is posterior-Riccati regret minus gated-critic regret;
material harm means deterioration greater than $0.001$. Cross-fit is ungated,
and exact-assisted is a mechanism reference.}
\label{tab:scalable-shift}
\centering
\scriptsize
\setlength{\tabcolsep}{4.0pt}
\begin{tabular}{lrrrrrrrr}
\toprule
& \multicolumn{5}{c}{Mean exact target-policy regret $\downarrow$}
& \multicolumn{3}{c}{Gated deployment} \\
\cmidrule(lr){2-6}\cmidrule(lr){7-9}
Target regime & Riccati & Pessimistic FQI & Cross-fit & Gated critic
& Exact-assisted & $\Delta_{\mathrm{R}}\uparrow$ & Accept & Material harm \\
\midrule
Stable   & 0.01297 & 0.13378 & 0.00543 & 0.00836 & 0.00596
         & 0.00461 & 55\% & 0\% \\
Moderate & 0.03101 & 0.20482 & 0.01343 & 0.01723 & 0.00628
         & 0.01379 & 70\% & 0\% \\
Severe   & 0.03489 & 0.10798 & 0.01358 & 0.01631 & 0.00628
         & 0.01858 & 85\% & 10\% \\
\bottomrule
\end{tabular}
\end{table}

\subsection{Model Validation and Numerical Checks}
\label{sec:validation-checks}

Before comparing policies, we check the observed transition model, pathwise
XVA accounting, and Riccati implementation separately.

The one-step model is adequate for the four variables that directly drive the
overlay. During the 2021--2025 OOD period, its nominal 90\% intervals cover
$91.5\%$--$97.6\%$ of observations for carbon returns, sector returns,
financial stress, and liquidity. The short rate is retained as an auxiliary
market state, not as one of the climate channels evaluated here; we make no
empirical validation claim for its equation. A stable VAR correction reduces
the largest lag-one residual autocorrelation from
$0.280$ to $0.036$, but does not improve OOD fit. Its only role is to stop
one-step errors from repeating; structural equations determine the scenario
difference.

Public data validate only the observed climate-on branch; they do not identify
a climate-off history. The baseline therefore remains the
documented pre-MSR scenario in Section~\ref{sec:data-protocol}. Starting from
each observed state, the two branches receive the same innovations and differ
only in climate-sensitive transition terms. The results are conditional on
this construction, not estimates of the MSR's historical or causal effect.

We next test the implementation rather than its predictive fit. On common
bootstrap paths, the gross liability, inherited hedge, residual liability,
overlay, funding, and terminal closeout satisfy their accounting identities
path by path. The inherited inventory is fixed before intervention and is the
same in both branches, so its incremental climate P\&L is credited once before
overlay optimization. Every reported policy respects the trading constraints.

Finally, we test the solver where an exact answer is available. In the
Gaussian quadratic case, the backward Riccati recursion matches an independently
assembled full-horizon quadratic program: the maximum inventory gap is
$3.33\times10^{-16}$ and the objective gap is zero. Nonquadratic proportional
spreads raise the maximum inventory-path gap to $0.0265$;
inventory bounds give $0.0161$. Regime switching, default jumps, path
dependence, and nonlinear liquidity move the full problem further from the
benchmark. Riccati is therefore exact only for the stated Gaussian quadratic
case; elsewhere it is the parent policy, not the final solution.

\subsection{Residual Climate-XVA and Hedging Performance}
\label{sec:hedging-results}

The gross charge is not the optimization target for a new climate mandate.
For the carbon-heavy book it is $1.517$, but the frozen rate and CDS positions
already reduce the amount passed to the overlay to $0.906$. Starting from this
residual, the certified policy reaches $0.831$, only $0.010$ above the exact
finite-MDP floor of $0.821$. It captures $88.4\%$ of the value still available
after the inherited hedge has been credited. Figure~\ref{fig:headline-story}A
shows this sequence.

The reduction comes from carbon exposure rather than from adding instruments
indiscriminately. Reoptimizing with no overlay, EUA only, utilities only, and
both instruments gives charges of $0.906$, $0.821$, $0.906$, and $0.821$.
The full $0.085$ Shapley value is therefore assigned to EUA, while utilities
are never held on reachable oracle states. This finding belongs to the stated
book and transaction costs; it is not a general ranking of climate hedges.

We then ask whether the control gain comes from the residual construction or
merely from simulated data. Exact regret is the entropic objective of a fixed
policy minus the finite-environment optimum. Residual Dyna reaches mean regret
$0.00757$ after 30 updates and 6,000 gradient trajectories; observed replay
remains at $0.10863$ after 120 updates and 24,000 trajectories. The paired
difference is $0.10107$, with a 95\% interval of $[0.06391,0.14219]$.
Residual Dyna uses one quarter of the trajectories and lowers regret by
$93\%$.

Simulation alone does not explain the result. Model-based training from
scratch at 120 updates is still worse than residual Dyna at 30 updates by
$0.03822$ ($[0.01674,0.06164]$). The Riccati parent supplies the ordinary
hedge direction; Dyna learns the remaining effects of regime, basis,
liquidity, and constraints.

\subsection{Component Analysis and Adaptation}
\label{sec:components-adaptation}

The ablation first asks what makes the RL policy learn efficiently. Across 20
fresh datasets, learning the whole policy from scratch is the baseline.
Writing the actor as a correction to Riccati reduces regret by $0.03752$
with interval $[0.01776,0.05934]$, the largest resolved component effect. The
robust objective adds $0.000795$, with interval $[0.000145,0.001732]$. At this budget, the effects
of observed seeding and importance correction are not resolved, while priority
without importance correction increases regret by $0.000884$, with interval
$[0.000051,0.001991]$. Residual parameterization produces the main gain: RL
learns the error left by the analytical policy instead of relearning its hedge.

We next give the residual learner only 25 transitions after a stable, moderate,
or severe market shift. An exact-assisted reference establishes whether
adaptation is useful at all. Under moderate and severe shifts, its gated update
removes $78.7\%$ and $87.9\%$ of posterior-Riccati regret. With 250
transitions, posterior Riccati is already close to the exact optimum and the
gate accepts no update. This reference measures the available adaptation gain
and is not used for deployment.

The deployable test removes exact values from critic fitting, actor
distillation, selection, and gating. The critic is cross-fitted over eight
folds and four independently fitted world models; the minimum importance
effective sample size is $0.499$. Exact target values remain hidden until all
policies have been fixed on 20 unused seeds.

Figure~\ref{fig:headline-story}B separates model updating from RL adaptation.
The navy square is posterior Riccati after observing 25 target transitions;
the green circle adds the gated RL correction. Since lower regret is better,
their leftward distance is the value learned beyond the updated model. Regret
falls from $0.01297$ to $0.00836$ in the stable environment, from $0.03101$ to
$0.01723$ under moderate shift, and from $0.03489$ to $0.01631$ under severe
shift. The larger gains under shift occur where posterior Riccati remains
misspecified. The orange diamond is an exact-assisted reference, not a
deployable policy. Across the moderate and severe shifts, the gated critic
improves Riccati by $0.01618$ on average, has a paired 95\% bootstrap lower
bound of $0.00992$, and retains $60.7\%$ of the exact-assisted gain.

Table~\ref{tab:scalable-shift} shows why the residual form matters with only 25
transitions. It also reports an ungated cross-fitted critic and pessimistic
FQI. The gated critic improves on FQI by $0.13963$ on average. FQI must fit a
sparsely covered state--inventory grid; the residual critic learns around the
Riccati action.

The final gate decides whether that learned correction may replace its parent.
In Figure~\ref{fig:headline-story}B, ``Gate'' is the share of candidate updates
accepted on 20 evaluation seeds, and ``harm'' is the share for which the final
policy is worse than Riccati by more than $0.001$. Acceptance is $55\%$,
$70\%$, and $85\%$ under stable, moderate, and severe conditions; material
harm is $0\%$, $0\%$, and $10\%$. With zero tolerance, five severe-shift seeds
deteriorate, but three changes are below $0.001$. The gate controls material
deterioration at the stated tolerance; individual seeds may still worsen.

\section{Conclusion and Limitations}
\label{sec:conclusion}

Climate-Dyna formulates deep hedging for XVAs as a residual control problem.
The inherited XVA hedge is credited first, and model-based RL trades only an
incremental climate overlay against the liability that remains. The optimized
cost under the largest feasible hedge universe is residual climate HVA.
Repeating the same optimization as instruments are added gives the
hedge-instrument discovery criterion. A finite-horizon Riccati policy anchors
the overlay; paired world-model Dyna learns its nonlinear correction, and
independent rollouts gate deployment.

The numerical evidence is a proof of concept, not a complete empirical
realization of the framework. Within the semi-synthetic EU ETS environment,
the experiments verify the pathwise accounting, recover the quadratic
benchmark, reduce control error, and test gated adaptation. The environment
instantiates transition, credit, sector-market, and liquidity channels, but
the climate-off law is scenario-defined and the short-rate OOD diagnostic is
weak. Exact regret is used only for evaluation, and the instrument attribution
is specific to the chosen book and costs. Physical risk, XCE term-sheet
calibration, richer and continuous hedge books, additional instrument groups,
causal policy estimation, and live-desk validation remain open.

\bibliographystyle{unsrt}
\bibliography{references}

\end{document}